\documentclass[aps,prb,twocolumn,superscriptaddress,noshowpacs]{revtex4}
\usepackage{color}
\usepackage{booktabs}
\usepackage{epsfig}
\usepackage[dvipdfmx]{hyperref}
\hypersetup{colorlinks=true,citecolor=blue,linkcolor=red}
\usepackage[all]{hypcap}
\usepackage{url}
\begin{document}
\newdimen\heavyrulewidth
\newdimen\lightrulewidth
\newdimen\cmidrulewidth
\newdimen\belowrulesep
\newdimen\belowbottomsep
\newdimen\aboverulesep
\newdimen\abovetopsep
\newdimen\defaultaddspace
\heavyrulewidth=.08em
\lightrulewidth=.05em
\belowrulesep=.65ex
\belowbottomsep=.0pt
\aboverulesep=.4ex
\abovetopsep=.0pt
\defaultaddspace=.5em

\title{ Local-Field Effects in Silicon Nanoclusters }

\author{\firstname{Roberto} \surname{Guerra}}
\affiliation{Dipartimento di Scienze e Metodi dell'Ingegneria and Centro Interdipartimentale En\&Tech , Universit\`a di Modena e Reggio Emilia, via Amendola 2 Pad. Morselli, I-42122 Reggio Emilia, Italy.}
\affiliation{Centro S3, CNR-Istituto di Nanoscienze, via Campi 213/A I-41100 Modena Italy.}

\author{\firstname{Margherita} \surname{Marsili}}
\author{\firstname{Olivia} \surname{Pulci}}
\affiliation{European Theoretical Spectroscopy Facility (ETSF) and CNR-INFM, Dept. of Physics, Universit\`a di Roma "Tor Vergata" Via della Ricerca Scientifica 1, I-00133 Roma, Italy.}

\author{\firstname{Stefano} \surname{Ossicini}}
\affiliation{Dipartimento di Scienze e Metodi dell'Ingegneria and Centro Interdipartimentale En\&Tech , Universit\`a di Modena e Reggio Emilia, via Amendola 2 Pad. Morselli, I-42122 Reggio Emilia, Italy.}
\affiliation{Centro S3, CNR-Istituto di Nanoscienze, via Campi 213/A I-41100 Modena Italy.}

\begin{abstract}
The effect of the local fields on the absorption spectra of silicon nanoclusters (NCs), freestanding or embedded in SiO$_2$, is investigated in the DFT-RPA framework for different size and amorphization of the samples. We show that local field effects have a great influence on the optical absorption of the NCs. Their effect can be described by two separate contributions, both arising from polarization effects at the NC interface. First, local fields produce a reduction of the absorption that is stronger in the low energy limit. This contribution is a direct consequence of the screening induced by polarization effects on the incoming field. Secondly, local fields cause a blue-shift on the main absorption peak that has been explained in terms of perturbation of the absorption resonance conditions. Both contributions do not depend neither on the NC diameter nor on its amorphization degree, while showing a high sensitivity to the environment enclosing the NCs.
\end{abstract}
\maketitle

\section{Introduction}\label{sec.intro}
\begin{table*}[t!]
  \centering
    \begin{tabular}[t]{c @{\hspace{0.6cm}} c @{\hspace{0.3cm}} c @{\hspace{0.3cm}} c @{\hspace{0.3cm}} c @{\hspace{0.3cm}} c @{\hspace{0.3cm}} c @{\hspace{0.3cm}} c}
      \toprule
        Structure         & NC-Si & core-Si & Si-centered & interface-O & bridge-bonds & $d$ (nm) & $V_s$ (nm$^3$)\\
      \midrule
	Si$_{10}$/SiO$_2$     & 10      & 0     & No          & 16          & 0            & 0.6      & 2.65\\
	Si$_{17}$/SiO$_2$     & 17      & 5     & Yes         & 36          & 0            & 0.8      & 2.61\\
	Si$_{32}$/SiO$_2$     & 32      & 12    & No          & 56          & 0            & 1.0      & 8.72\\
      \midrule
	a-Si$_{10}$/a-SiO$_2$ & 10      & 1     & Yes         & 20          & 0            & 0.6      & 2.61\\
	a-Si$_{17}$/a-SiO$_2$ & 17      & 5     & Yes         & 33          & 3            & 0.8      & 2.49\\
	a-Si$_{32}$/a-SiO$_2$ & 32      & 7     & No          & 45          & 3            & 1.0      & 8.67\\
      \bottomrule
    \end{tabular}
    \caption{\it Structural characteristics for the crystalline embedded NCs (top set) and amorphous embedded NCs (bottom set). For each structure are reported, respectively: number of Si atoms forming the NC (NC-Si), number of Si atoms forming the NC and not bonded with oxygens (core-Si), whether the NC is centered or not on one silicon (Si-centered), number of oxygens bonded to the NC (interface-O), number of oxygens bridging two NC-Si (bridge-bonds), average diameter $d$, supercell volume $V_s$.} \label{table1}
\end{table*}

The indirect nature of the energy band-gap in silicon has always been the major obstacle for its employment in light emitting devices, since the momentum conservation requires additional mechanisms involved in the recombination process (e.g. electron-phonon interaction), that occur with low probability, hence producing very poor emitting rates.
In the last decade the discovery of efficient visible photoluminescence (PL) and optical gain from silicon nanoclusters (Si-NCs) has demonstrated the possibility to overcome the indirect band gap of silicon by exploiting the quantistic behaviour of the matter at the nanoscale.\cite{pavesi, pavlockwood}\\
\noindent Theoretically, the optical emission has been attributed to transitions between states localized inside the nanocrystal [as a consequence of the so-called quantum confinement (QC) effect],\cite{klimov, moskalenko, hill_whaley, derr_rosei} or between defect states.\cite{dovrat, kanemitsu, iwayama, averboukh, koponen, borczyskowski} While there is still some debate on which of the above mechanisms primarily determines the emission energy, some recent works have proposed that a concomitance of both mechanisms is always present, favouring one or the other depending on the structural conditions.\cite{wolkin, allan, hao_green, gourbilleau, zhou, lin_chen, rolver, PRB2, daldosso, prakash} In the attempt of explaining the observations, it was suggested that for NC diameters above a certain threshold (of about 3 nm) the emission peak simply follows the QC model, while interface states would assume a crucial role only for small-sized NCs. Anyway, such consclusion was still unsatisfactory in many case. More recently, in a brilliant experiment Godefroo et al.\cite{godefroo} solved the puzzle demonstrating, by using magnetic fields to tune the QC and UV lasers to induce defects, that it is possible to reversibly control the origin of the PL by introducing or removing defects in a single sample: in the former case the PL originates from defects while in the latter case it originates from QC.\\

Previous works already highlighted the dramatic sensitivity of the opto-electronical properties to the Si/SiO$_2$ interface configuration, especially for very small NCs ($d \lesssim 1$ nm), where a large proportion of the atoms is localized at the interface. In the latter case, several NC characteristics such as passivation, symmetry, and strain, considerably concur for the determination of the final opto-electronic response, producing sensible deviations from the simple QC model.\cite{PRB2}
Moreover, many PL experiments demonstrated that only a very small fraction of the NCs in the samples contributes to the observed PL, enforcing the idea that precise structural conditions are required in order to achieve high emission rates.\cite{miura,credo,belomoin,dobrovolskas}
Finally, recent calculations reported especially high optical yields for small NCs,\cite{PRB3}, enhancing the weight held by their contribution in real samples. It is therefore clear the importance of understanding the factors that, at these sizes, contribute to enhance (or reduce) the NC optical response.

\noindent Embedding Si-NCs in wide band-gap insulators is one way to obtain a strong QC. Si-NCs embedded in a silica matrix have been obtained by several techniques as ion implantation,\cite{iwayama,daldosso,prakash,brongersma} chemical vapour deposition,\cite{hernandez,rolver,godefroo,gardelis} laser pyrolysis,\cite{delerue,kanemitsu} electron beam lithography,\cite{sychugov_linnros} sputtering,\cite{averboukh,antonova} and some others.\\
\noindent Experimentally, several factors contribute to make the interpretation of measurements on these systems a difficult task. For instance, samples show some dispersion in the NC size, that is difficult to be controlled. In this case it is possible that the observed quantity does not correspond exactly to the mean size but instead to the most responsive NCs.\cite{credo} Moreover, NCs synthesized by different techniques often show different properties in size, shape and in the interface structure. Finally, in solid nanocrystal arrays some collective effects caused by electron, photon, and phonon transfer between the NCs can strongly influence the electron dynamics in comparison with the case of isolated NCs.\cite{iwayama}
In practice, all the conditions remarked above lead to measurements of collective quantities, making the identification of the most active configurations at the experimental level a non trivial task.\\
\noindent From the theoretical side, the possibility of atomically manipulating the structures and of associating the selected configuration to the calculated response allows in principle to elucidate some of the points raised above. However, it must be taken into account that an accurate characterization of the electronical properties requires a full {\it ab-initio} approach, limiting the systems size to a few thousands of atoms in the case of density functional theory (DFT) methods. In addition, the calculation of realistic optical absorption or emission spectra, involving excited states, requires refined treatments that dramatically increase the computational effort, furtherly reducing the maximum manageable system size.\\

\noindent In this work we present DFT calculations in the local density approximation (LDA) of the ground-state electronic configuration of Si-NCs of different size embedded in a SiO$_2$ matrix, both in the amorphous and in the crystalline phase. The reason for taking into account amorphous NCs is based on the fact that real samples are always characterized by a certain amount of amorphization, in particular for NCs of small diameter.\cite{rinnert,gourbilleau,veprek} The absorption spectra, represented by the imaginary part of the dielectric function, are then evaluated within the random-phase approximation (RPA), with and without the inclusion of local field effects (LFE).\\
\noindent It is well known that the DFT-LDA severely underestimates the band gaps for semiconductors and insulators. A correction to the fundamental band gap is usually obtained by calculating the separate electron and hole quasiparticle energies via the GW method.\cite{review_onida_reining_rubio} In this method the self-energy $\Sigma$ is expanded in terms of the single particle Green function $G$ and the screened Coulomb interaction $W$, and at the first order is truncated to the first term $\Sigma\simeq iGW$. The knowledge of the quasiparticle energies, however, is still not sufficient to correctly describe a process in which electron-hole pairs are created such as the light absorption process. If the electron and the hole, created during the absorption process, are considered as independent quasiparticles, the structures of the absorption spectrum are located at the differences between the corresponding single-quasiparticles excitation levels. However, the attractive interaction between the positively and negatively charged quasiparticles can lead to a strong shift of the peak positions as well as to distortions of the spectral lineshape, known as excitonic effects. Within the many-body perturbation theory (MBPT) framework such interaction is taken into account by the solution of the Bethe-Salpeter equation for the polarizability.\cite{MBPT}\\
\noindent An alternative approach to MBPT for the computation of neutral excitations is represented by time-dependent DFT (TDDFT).\cite{TDDFT} TDDFT is expected to be more efficient than the MBPT-based approach, however many conceptual and computational problems remains still unsolved preventing its application to complex systems.\cite{review_onida_reining_rubio} Moreover, a recent comparison of the two techniques applied to Si-NCs revealed that TDDFT does not take into account correctly the screened Coulomb interaction, also for small NCs.\cite{ramos_paier_kresse_bechstedt}\\
\noindent Even if complex MBPT treatments should be invoked to include self-energy and excitonic effects, previous many-body calculations on Si-NCs reported fundamental gaps\cite{delerue_lannoo_allan} and absorption spectra\cite{PRB1,ramos_paier_kresse_bechstedt, gatti_onida} very close to the independent-particle calculated ones when LFE are neglected. In Ref. \onlinecite{PRB1} we verified this statement for the Si$_{10}$ and a-Si$_{10}$ embedded NCs, showing that self-energy corrections (calculated through the GW method) and electron-hole Coulomb corrections (calculated through the Bethe-Salpeter equation) nearly cancel out each other (with a total correction to the gap smaller than 0.2 eV). These considerations justify our choice of DFT-LDA for the calculation of the optical properties of larger clusters, allowing a good compromise between results accuracy and computational effort.\\

\section{Structures and Method}\label{sec.method}
\noindent The crystalline embedded structures have been obtained from a betacristobalite cubic supercell by removing all the oxygens included in a cutoff-sphere, whose radius determines the size of the NC. By centering the cutoff-sphere on one silicon or in an interstitial position it is possible to obtain structures with different symmetries. To guarantee a proper shielding of the strain arising from the difference in the silicon/silica lattice spacing, we preserved a separation of about 1 nm between the NCs replica.\\
\begin{figure*}[t]
  \centerline{\includegraphics[draft=false,height=18cm,angle=270]{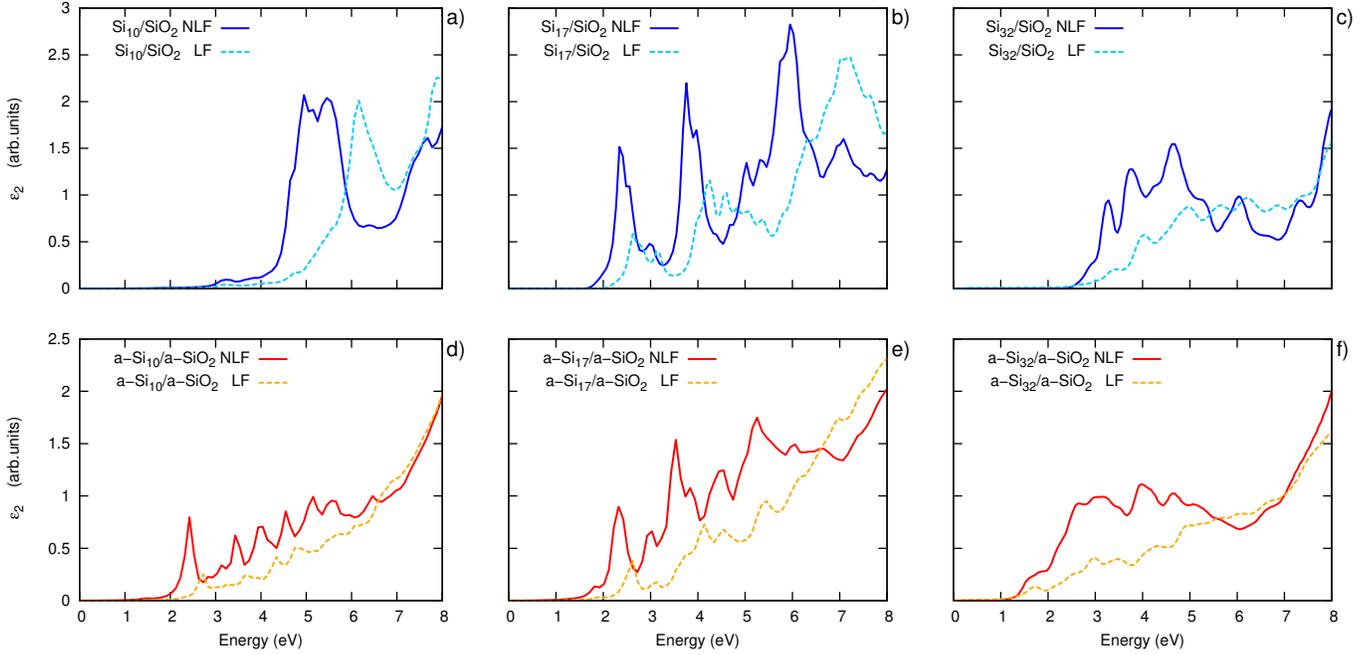}}
  \caption{\it (color online) Absorption spectra for the crystalline (top figures) and amorphous (bottom figures) sets of embedded NCs, with (dashed line) and without (solid line) the contribution of local fields.}\label{fig.LF-vs-NLF}
\end{figure*}
\noindent The glass models have been generated using classical molecular dynamics simulations of quenching from a melt, followed by \textit{ab-initio} relaxations (See Ref. \onlinecite{PRB1} for further details). The amorphous dot structures have been obtained, like in the case of the crystalline systems, by applying the cutoff sphere on the glass supercells.\cite{notaamorph} It is worth to stress that also in the crystalline case the embedding matrix looses its symmetry after the inclusion of the NC, due to the metastable nature of the betacristobalite.\cite{PRB1}\\
\noindent The relaxation of all the structures have been performed using the SIESTA code\cite{siesta1,siesta2} with a DZP basis set (double-$\zeta$ basis plus polarization orbitals) and Troullier-Martins pseudopotentials with non-linear core corrections. A cutoff of 250 Ry on the density and no additional external pressure or stress have been applied. Atomic positions and cell parameters have been left totally free to move.\\
\noindent Following the procedure described above we have built three crystalline embedded nanostructures, Si$_{10}$, Si$_{17}$, Si$_{32}$, and their respective amorphous counterparts, a-Si$_{10}$, a-Si$_{17}$, a-Si$_{32}$. The structural characteristics of all the systems are reported in Table \ref{table1}.\\
\noindent In order to investigate the role of the embedding medium on the LFE we have also built the freestanding counterparts of the crystalline and amorphous Si$_{32}$ NCs by extracting the Si-NCs from the silica, including or not the first shell of interface oxygens, and then by passivating the dangling bonds with hydrogen. In this way we have obtained four freestanding NCs: Si$_{32}$-(OH)$_{56}$, Si$_{32}$-H$_{56}$, a-Si$_{32}$-O$_{45}$-H$_{42}$, a-Si$_{32}$-H$_{48}$. For such structures, in order to preserve the strain induced by the silica matrix on the NCs,\cite{PRB2} we performed a structural relaxation on the sole hydrogens, while holding the position of the other atoms. The structural properties of the freestanding structures are therefore identical to the embedded ones.\\
\noindent For each structure we have calculated the eigenvalues and eigenfunctions of the Kohn-Sham Hamiltonian using the ESPRESSO package. \cite{espresso} Calculations have been performed using norm-conserving pseudopotentials within the Local Density Approximation (LDA) with a Ceperley-Alder exchange-correlation potential, as parametrized by Perdew-Zunger. An energy cutoff of 60 Ry on the plane wave basis set has been considered.\\
\noindent Once the ground-state geometry has been found, the absorption spectra have been computed at the DFT-RPA level with and without the inclusion of LFE. The absorption spectrum, given by the imaginary part of the macroscopic dielectric function $\epsilon_M(\omega)$, is connected to the inverse of the microscopic dielectric function $\epsilon^{-1}_{GG'}(q,\omega)$ through the so called `macroscopic average':\cite{adler}
\begin{equation}
\epsilon_M(\omega)=\lim_{q\rightarrow0}\frac{1}{\epsilon^{-1}_{00}(q,\omega)}.
\label{eq:macraverage}
\end{equation}
When LFE are neglected at the RPA level, $\epsilon_M(\omega)=\lim_{q\rightarrow0}\epsilon_{00}(q,\omega)=1-\lim_{q\rightarrow0}v(q)P^0_{00}$, where $v(q)$ is the Coulomb interaction and $P^0$ is the irreducible RPA polarizability. This procedure is in fact exact in the case of an homogeneous system for which the off-diagonal terms of $\epsilon^{-1}_{GG'}(q,\omega)$ are null.
On the other hand, when local fields (LF) are included the quantity  $\epsilon^{-1}_{00}(q,\omega)$ must be accessed. Very briefly, $\epsilon^{-1}$ is linked to the reducible polarizability $\chi$ by the relation $\epsilon^{-1}=1+v\chi$. At the RPA level we have that $\chi=P^0+P^0v\chi$. Hence by calculating $P^0=-iG^0G^0$ with $G^0$ single particle Green function, we can obtain $\chi$ and $\epsilon^{-1}$.\\

\section{Results}\label{sec.results}
\noindent It is widely known that LFE assume a crucial role for systems characterized by strong charge inhomegeneities. Instead, for ordered systems like bulk-silicon and betacristobalite, LFE tend to vanish out.\cite{louie-chelikowsky-cohen} Besides, the same rule applies also for completely amorphized systems, like silica-glasses, that at last behave as homogeneous materials.\cite{LFglass} In the case of Si/SiO$_2$ heterostructures, the inhomogeneity is represented by the interface that the NC forms with the surrounding silica, and it is therefore important to investigate the role of LF for systems with different interface conditions.

\noindent In Figure \ref{fig.LF-vs-NLF} we report the DFT-RPA absorption spectra with and without the LFE contribution, for the crystalline and amorphous Si$_{10}$, Si$_{17}$, and Si$_{32}$ embedded NCs.\cite{pssb} We note that the microscopic field fluctuations produce important screening effects on the spectra of all the systems, with a damping of the absorption that is more effective at low energies. Instead, at energies larger than the optical gap of the embedding medium ($\sim5.5$ eV for DFT-LDA) the NLF and LF spectra present much similar profiles, suggesting that in this energy regime the absorption is completely due to pure SiO$_2$ states, for which the LFE are absent. We also observe a blue-shift on the main absorption peaks, in particular for the smaller systems in the crystalline phase. The origin of such a shift is not clear at this point. We can elaborate on this aspect with the aid of a modified Lorentz oscillator model, adapted in order to include the interface polarization effects (IPE):
\begin{equation}
	\ddot{x}+2\xi\omega_0\dot{x}+\omega_0^2x=E_0\sin(\omega t)-\alpha x ~~~~~. \label{eq.moto}
\end{equation}
In Eq. \ref{eq.moto} the oscillator coordinate, $x$, corresponds to the electronic density displacement, $\omega_0$ is the undamped frequency, $\xi$ is a constant (usually called {\it damping ratio}) associated to self-relaxation processes, $E_0\sin(\omega t)$ is the external field, and $\alpha x$ is the field produced by the interface polarization, that contrasts the external field. The solution is
\begin{equation}
	x(t)=E_0\left[\left(2\omega\omega_0\xi\right)^2+\left(\omega_0^2+\alpha-\omega ^2\right)^2\right]^{-\frac{1}{2}} \sin\left[\omega t+\phi\right] ~~~~,  \label{eq.solution}
\end{equation}
where $\phi=\arctan\left[{2\omega\omega_0\xi}/{(\omega^2-\omega_0^2-\alpha)}\right]$ is the phase-shift that determines the screening. The blue-shift appearing in the LFE spectra of Fig. \ref{fig.LF-vs-NLF} emerges in the model as a consequence of the additional term, $-\alpha x$, that changes the resonance frequency from $\omega_0$ to $\omega_0'=\sqrt{\omega_0^2+\alpha}$. In this picture, damping of the absorption and shifting of the resonance peak arise from the same physical quantity, and are therefore intimately connected.
\\It is interesting at this point to examine the response of the model in upper and lower outermosts of the driving field. At low frequencies, $\omega\ll\omega_0'$, the interface polarization oscillates in phase with the external field due to a small phase-shift $\phi$. The external field is therefore maximally screened in this regime. In principle, the same situation occurs at high frequencies, $\omega\gg\omega_0'$, where the phase-shift is a decreasing function approaching zero. Besides, the function of $\omega$ multiplying the sine in Eq. \ref{eq.solution} tends to zero in the limit of high frequencies, leading to vanishing LFE. At $\omega\sim\omega_0'$ the interface polarization oscillates $\pi/2$ out of phase from the external field that gets therefore anti-screened, leading to the formation of the (shifted) absorption peak.\\

\begin{figure}[b!]
  \centerline{\includegraphics[draft=false,height=8.5cm,angle=270]{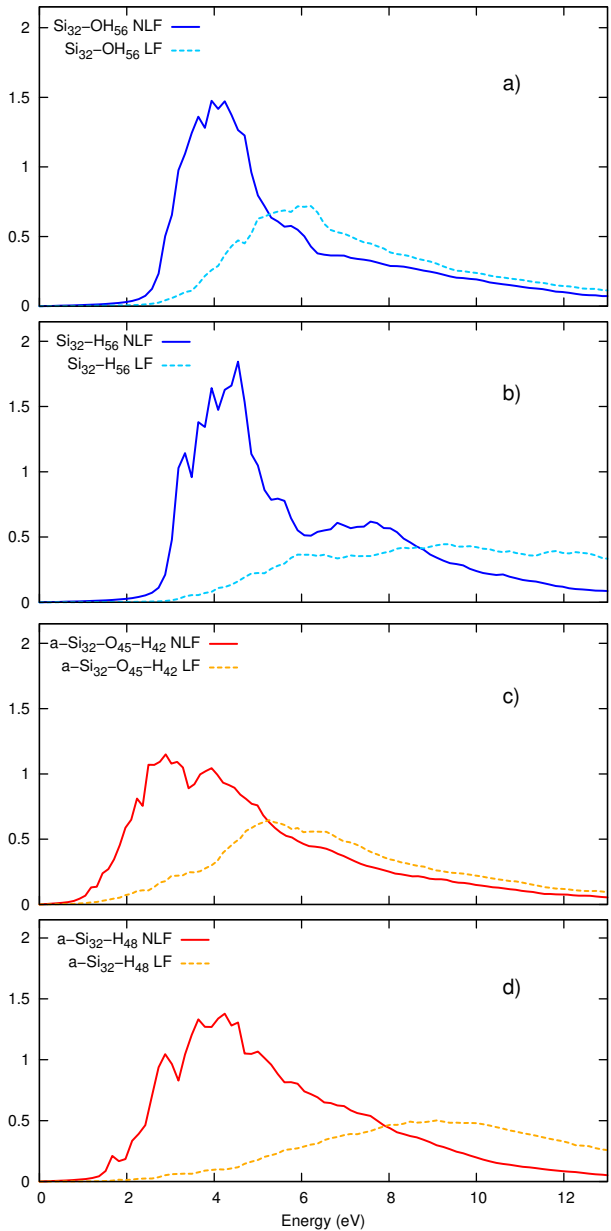}}
  \caption{\it (color online) Absorption spectra for the Si$_{32}$ freestanding NCs, with (dashed line) and without (solid line) the contribution of local fields: (a) crystalline OH-terminated, (b) crystalline H-terminated, (c) amorphous OH-terminated, (d) amorphous H-terminated. The spectra are in arbitrary units. }\label{fig.LF-vs-NLF_suspended}
\end{figure}

\noindent It is worth to note that LFE have a strong influence at every NC size. This result is in agreement with previous works reporting high LFE on very large nanostructures, suggesting that they arise almost entirely by classical effects.\cite{Bruneval2005, Trani2007} In addition, we observe similar trends of the spectra of the two largest NCs (Si$_{32}$ and a-Si$_{32}$). This supports the idea that for large NCs the response should depend mainly on the IPE and not on the amorphization degree, nor on the particular geometry of the Si/SiO$_2$ interface. Finally, differently from Ref. \onlinecite{bulutay2007}, we observe similar absorption magnitudes for different NCs size (especially in the more realistic case of amorphous NCs), in agreement with recent experimental observations.\cite{mirabella}
\\The comparison of the LF spectra of Figs. \ref{fig.LF-vs-NLF}c,\ref{fig.LF-vs-NLF}f with experimental measurements on 1nm-sized Si/SiO$_2$ NCs show a nice match of the absorption profile and of the maximum absorption energy located at about 6 eV.\cite{gallas}

\noindent In order to discuss the role of the embedding medium we have considered also the case of freestanding NCs.
In Ref.~\onlinecite{PRB1} we showed that, while the freestanding hydrogenated NCs present a larger band-gap and miss oxygen-related states at the band-edge, the NC+interface system is able (when the strain induced by the embedding matrix is preserved) to nicely reproduce the characteristics of the full NC+SiO$_2$ system.
\\In Figure \ref{fig.LF-vs-NLF_suspended} we report the absorption spectra for the set of freestanding Si$_{32}$ NCs, with and without the inclusion of the LFE. First of all we note that the NLF results confirm the point remarked above: the NLF spectra of the OH-terminated NCs (Figs. \ref{fig.LF-vs-NLF_suspended}a and \ref{fig.LF-vs-NLF_suspended}c) well match those of the corresponding embedded NCs in the 0--6 eV range, while both the NLF spectra of the hydrogenated NCs (Figs. \ref{fig.LF-vs-NLF_suspended}b and \ref{fig.LF-vs-NLF_suspended}d) present some modifications and a blue-shift of about 0.5 eV on the main peak.

\begin{figure*}[t!]
  \centerline{\includegraphics[draft=false,height=18cm,angle=270]{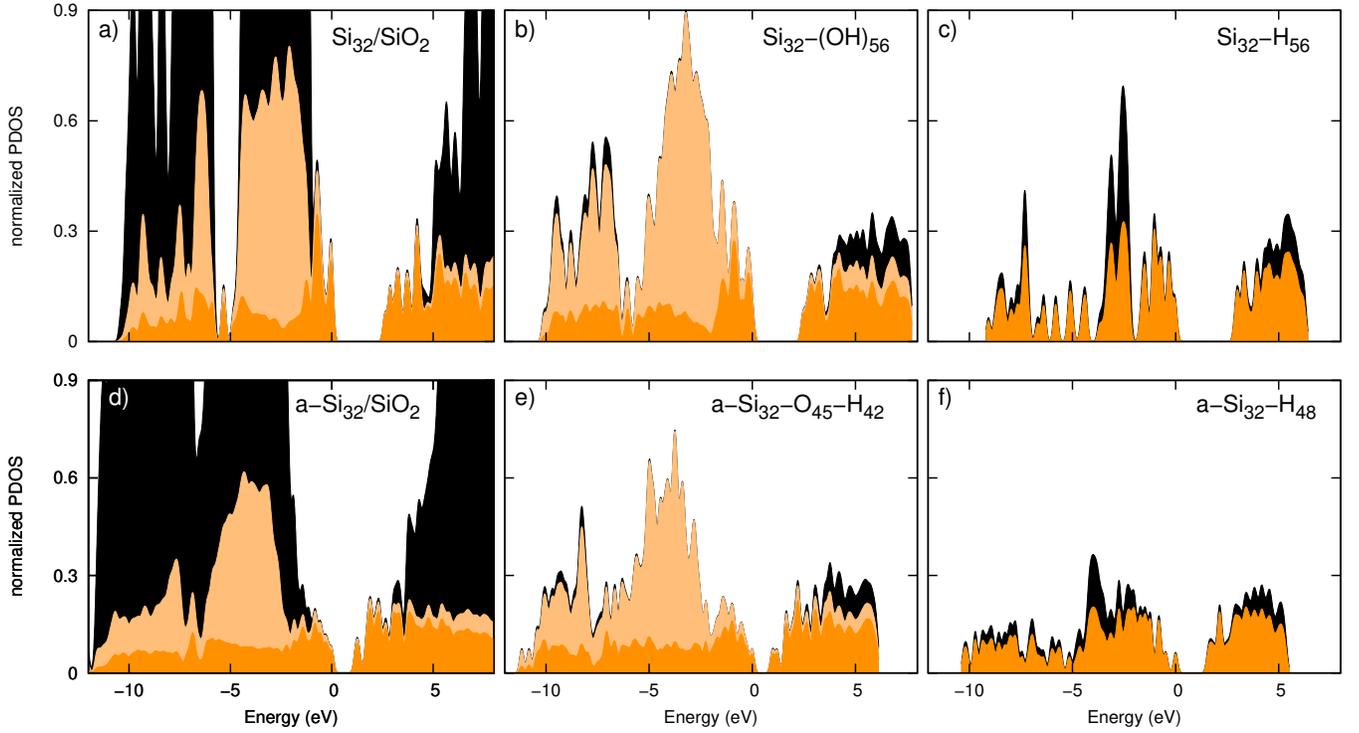}}
  \caption{\it (color online) Projected-DOS for the Si$_{32}$ NCs: (a) cristalline embedded,  (b) crystalline OH-terminated, (c) crystalline H-terminated, (d) amorphous embedded, (e) amorphous OH-terminated, (f) amorphous H-terminated. The projections on the NC silicons, NC+interface, and all the atoms are highlighted in orange (gray), light-orange (light-gray), and black, respectively. The PDOS are normalized following the procedure described in the text. A gaussian broadening of 0.1 eV has been applied to the data.}\label{fig.pdos}
\end{figure*}

\noindent Also the LF spectra of the NC+interface systems match nicely those of the corresponding full systems, evidencing a strong screening in the 0--4 eV range and a blue-shift of about 2 eV on the main peak. These spectra are nicely comparable with those of Ref. \onlinecite{gallas} obtained by an effective-medium approach from experimental measurements on embedded NCs. It is worth to stress that in our samples the crystallinity degree and the amorphization degree are intentionally maximized. It is therefore reasonable to observe spectra of real 1nm-sized NCs lying in-between those of Figs. \ref{fig.LF-vs-NLF_suspended}a and \ref{fig.LF-vs-NLF_suspended}c.\\
In the case of hydrogenated NCs the LFE become of dramatic proportions, with severe screening effects and a shifting of the main peak of about 5 eV, both ascribable (by the model discussed above) to a faster response of the iterface polarization. At last it is worth noticing that, as already evidenced for the embedded case, the introduction of LFE produces spectra that are poorly dependent on the amorphization degree. This consideration, together with the fact that in real samples the smaller NCs tend to remain amorphous,\cite{rinnert,gourbilleau,veprek} is positive in photovoltaics applications, where an efficient harvesting of the sunlight spectrum is required. The small variation of the response on the amorphization degree is in this case an advantage, because any variation could rule the smallest NCs out of the absorption process.\\

\noindent In the last part of this section we investigate on the different contributions of Si-NC, interface, and enclosing environment, to the absorption spectra.
\\In Figure \ref{fig.pdos} we report the density of states (DOS) projected on the NC silicons, NC+interface, and all the atoms, of the embedded and freestanding Si$_{32}$ systems. All the projected DOS (PDOS) have been normalized following the constraint
\begin{equation}
 \int_{-\infty}^{E_F} PDOS(E)dE = 1 ~~~, \label{eq.DOS_normalization}
\end{equation}
where $PDOS(E)$ is the DOS projected on the NC silicons, and $E_F$ is the Fermi energy located half-between the energy band gap.
\\In the case of embedded NCs (Figs. \ref{fig.pdos}a, \ref{fig.pdos}d) the contribution due to the interface-oxygens is very small concerning the conduction band while it assumes an important role in the valence band, especially for energies below -1 eV (-2 eV in the amorphous case). It follows that, at least at the NLF level, the absorption below $\sim$3.4 eV ($\sim$2.9 eV in the amorphous case) is mostly due to states localized on the NC, while at higher energies also interface-to-NC transitions occur. As expected, the PDOS for the OH-terminated NCs (Figs. \ref{fig.pdos}b, \ref{fig.pdos}e) closely reproduce those of the corresponding embedded counterparts, confirming the idea that the NC+interface system is non-interacting with the remaining atoms.\cite{PRB1,pssb} Instead, in the hydrogenated case the PDOS of the NCs present a raising in the -2~--~-5 eV region with respect to the other systems, that is responsible for the increased absorption in the 6--8 eV (see NLF spectra of Figs. \ref{fig.LF-vs-NLF_suspended}b, \ref{fig.LF-vs-NLF_suspended}d).

\section{Conclusions}\label{sec.conclusions}
\noindent Local field effects (LFE) have a great influence on the optical absorption of silicon nanoclusters (Si-NCs), and must be included for a realistic description of the optical response. Their effect arise from the polarizations occurring at the NC interface (IPE), and they are therefore particularly sensitive to the details of the NC termination.
\\The IPE depend on the energy of the incoming field. At low energies the incoming field and the induced polarizations are in-phase, resulting in a screened effective field and a subsequent damping of the absorption. Contrary, at high energies the polarizations are not able to follow the fast oscillations of the incoming field, and LFE tend to vanish out. Besides, at energies for which the incoming field and the induced polarization are opportunely dephased, the former becomes anti-screened, leading to a maximized absorption. Therefore, the final (corrected) spectrum appears reduced and blue-shifted, by magnitudes that depend on the interface polarizability conditions. In presence of interface oxygens, as for freestanding like as for embedded NCs, the LFE produce a severe reduction of the absorption up to 4 eV, and a blue-shift on the main peak settling around 2 eV. In the case of hydrogenated NCs the LFE are of dramatic proportions, with a severe damping of the absorption in the 0--4 eV energy range, and a blue-shift attaining an impressive value of about 5 eV. Interestingly, in all the case considered, while the uncorrected spectra show an important sensitivity to the structural configuration of the NCs (amorphization), LFE tend to smooth out such differences in favor of a more consistent description of the absorption characteristic. This result suggests the possibility of relying on simpler methods for the evaluation of the LFE, like the effective medium theories as already suggested for other silicon nanostructures.\cite{Bruneval2005} Such a simplification would be especially convenient when considering larger NCs, requiring to date prohibitive computational efforts with the current model.\\

\section*{Acknowledgements}
\noindent This work is supported by PRIN2007 and by Ministero degli Affari Esteri, Direzione Generale per la Promozione e la Cooperazione Culturale. R.G. acknowledge financial support from Fondazione Cassa di Risparmio di Modena under the project ``Progettazione di materiali nanostrutturati semiconduttori per la fotonica, l'energia rinnovabile e l'ambiente''. We acknowledge CINECA CPU time granted by INFM ``Progetto Calcolo Parallelo'' and CASPUR Standard HCP Grant 2010. The research leading to these results has received funding from the European Community's Seventh Framework Programme (FP7/2007-2013) under grant agreement no. 211956.


\begin{thebibliography}{}
\bibitem{pavesi} L. Pavesi, L. Dal Negro, C. Mazzoleni, G. Franz\`o, F. Priolo, Nature \textbf{408}, 440 (2000).
\bibitem{pavlockwood} L. Pavesi and D.J. Lockwood, Eds., Silicon Photonics, Springer, Berlin, 2004.
\bibitem{klimov} M. Sykora, L. Mangolini, R. D. Schaller, U. Kortshagen, D. Jurbergs, V. I. Klimov, Phys. Rev. Lett. {\bf 100}, 067401 (2008).
\bibitem{moskalenko} A. S. Moskalenko, J. Berakdar, A. A. Prokofiev, I. N. Yassievich, Phys. Rev. B {\bf 76}, 085427 (2007).
\bibitem{hill_whaley} N. A. Hill, K. B. Whaley, Phys. Rev. Lett. {\bf 75}, 1130 (1995).
\bibitem{derr_rosei} J. Derr, K. Dunn, D. Riabinina, F. Martin, M. Chacker, F. Rosei, Physica E {\bf 41} 668-670 (2009).
\bibitem{dovrat} M. Dovrat, Y. Shalibo, N. Arad, I. Popov, S.-T. Lee, A. Sa'ar, Phys. Rev. B {\bf 79}, 125306 (2009).
\bibitem{kanemitsu} Y. Kanemitsu, Thin Solid Films {\bf 276}, 44-46 (1996).
\bibitem{iwayama} T. Shimitsu-Iwayama, T. Hama, D. E. Hole, I. W. Boyd, Solid-State Electronics {\bf 45}, 1487-1494 (2001).
\bibitem{averboukh} B. Averboukh, R. Huber, K. W. Cheah, Y. R. Shen, G. G. Qin, Z. C. Ma, W. H. Zong, J. Appl. Phys {\bf 92}, 3564-3568 (2002).
\bibitem{koponen} L. Koponen, L. O. Tunturivuori, M. J. Puska, R. M. Nieminen, Phys. Rev. B {\bf 79}, 235332 (2009).
\bibitem{borczyskowski} J. Martin, F. Cichos, F. Huisken, C. von Borczyskowski, Nano Letters {\bf 8}, 656 (2008).
\bibitem{wolkin} M. V. Wolkin, J. Jorne, P. M. Fauchet, G. Allan, C. Delerue, Phys. Rev. Lett. {\bf 82}, 000197 (1999).
\bibitem{allan} G. Allan, C. Delerue, M. Lannoo, Phys. Rev. Lett. {\bf 76}, 2961 (1996).
\bibitem{hao_green} X. J. Hao, A. P. Podhorodecki, Y. S. Shen, G. Zatryb, J. Misiewicz, M. A. Green, Nanotechnology {\bf 20}, 485703 (2009).
\bibitem{gourbilleau} F. Gourbilleau, C. Ternon, D. Maestre, O. Palais, C. Dofour, J. Appl. Phys. {\bf 106}, 013501 (2009).
\bibitem{zhou} Z. Zhou, L. Brus, R. Friesner, Nano Lett. {\bf 3}, 163 (2003).
\bibitem{lin_chen} S.-W. Lin, D.-H. Chen, Small {\bf 5}, 72-76 (2009).
\bibitem{rolver} R. Rolver, M. F\"{o}rst, O. Winkler, B. Spangenberg, H. Kurz, J. Vac. Sci. Technol. A {\bf 24}, 141-145 (2006).
\bibitem{PRB2} R. Guerra, E. Degoli, S. Ossicini, Phys. Rev. B {\bf 80}, 155332 (2009).
\bibitem{daldosso} N. Daldosso, M. Luppi, S. Ossicini, E. Degoli, R. Magri, G. Dalba, P. Fornasini, R. Grisenti, F. Rocca, L. Pavesi, S. Boninelli, F. Priolo, C. Spinella, F. Iacona, Phys. Rev. B {\bf 68}, 085327 (2003).
\bibitem{prakash} G. V. Prakash, N. Daldosso, E. Degoli, F. Iacona, M. Cazzinelli, F. Rocca, Z. Gaburro, P. Dalba, E. C. Moreira, D. Pacifici, G. Franz\`o, F. Priolo, C. Arcangeli, A. B. Filonov, S. Ossicini, L. Pavesi, Journal of Nanoscience and Nanotechnology {\bf 1}, 159-168 (2001).
\bibitem{godefroo} S. Godefroo, M. Hayne, M. Jivanescu, A. Stesmans, M. Zacharias, O. I. Lebedev, G. V. Tendeloo, V. V. Moshchalkov, Nature Nanotech. {\bf 3}, 174 (2008).
\bibitem{miura} S. Miura, T. Nakamura, M. Fujii, M. Inui, Shinji Hayashi, Phys. Rev. B {\bf 73}, 245333 (2006). 
\bibitem{credo} G. M. Credo, M. D. Mason, S. K. Buratto, Appl. Phys. Lett. {\bf 74}, 1978 (1999).
\bibitem{belomoin} G. Belomoin, J. Therrien, A. Smith, S. Rao, R. Twesten, S. Chaieb, M. H. Nayfeh, L. Wagner, L. Mitas, Appl. Phys. Lett. {\bf 80}, 841 (2002).
\bibitem{dobrovolskas} D. Dobrovolskas, J. Mickevi\v{c}ius, G. Tamulaitis, V. Reipa, J. Phys. Chem. of Solids {\bf} 70, 439 (2009).
\bibitem{PRB3} R. Guerra, S. Ossicini, Phys. Rev. B {\bf 81}, 245307 (2010).
\bibitem{brongersma} M. L. Brongersma, P. G. Kik, A. Polman, Appl. Phys. Lett. {\bf 76}, 351-353 (2000).
\bibitem{hernandez} A. V. Hernandez, T. V. Torchynska, Y. Matsumoto, S. J. Sandoval, M. Dybiec, S. Ostapenko, L. V. Shcherbina, Microelectronics Journal {\bf 36}, 510-513 (2005).
\bibitem{gardelis} S. Gardelis, A. G. Nassiopoulou, N. Vouroutzis, N. Frangis, J. Appl. Phys. {\bf 105}, 113509 (2009).
\bibitem{delerue} C. Delerue, G. Allan, C. Reynaud, O. Guillois, G. Ledoux, F. Huisken, Phys. Rev. B {\bf 73}, 235318 (2006).
\bibitem{sychugov_linnros} I. Sychugov, R. Juhasz, J. Valenta, J. Linnros, Phys. Rev. Lett. {\bf 94}, 087405 (2005).
\bibitem{antonova} I. V. Antonova, M. Gulyaev, E. Savir, J. Jedrzejewski, I. Balberg, Phys. Rev. B  {\bf 77} 125318 (2008).
\bibitem{rinnert} H. Rinnert, M. Vergnat, A. Burneau, J. Appl. Phys. {\bf 89}, 237 (2001).
\bibitem{veprek} S. Veprek, Z. Iqbal, F. A. Sarott, Philos. Mag. B {\bf 45}, 137 (1982).
\bibitem{review_onida_reining_rubio} G. Onida, L. Reining, A. Rubio, Rev. Mod. Phys. {\bf 74}, 601 (2002).
\bibitem{MBPT} A. L. Fetter and J. D. Walecka, ``Quantum Theory of Many-Particle Systems'', New York:McGraw-Hill (1971).
\bibitem{TDDFT} E. Runge and E. K. U. Gross, Phys. Rev. Lett. {\bf 52} 997 (1984).
\bibitem{ramos_paier_kresse_bechstedt} L. E. Ramos, J. Paier, G. Kresse, F. Bechstedt, Phys. Rev. B {\bf 78}, 195423 (2008).
\bibitem{delerue_lannoo_allan} C. Delerue, M. Lannoo, G. Allan, Phys. Rev. Lett. {\bf 84}, 2457 (2000).
\bibitem{notaamorph} The reliability of such procedure has been tested by producing five different Si$_{10}$ amorphous NCs and by verifying that the resulting spectra were substantially equivalent.
\bibitem{PRB1} R. Guerra, I. Marri, R. Magri, L. Martin-Samos, O. Pulci, E. Degoli, S. Ossicini, Phys. Rev. B {\bf 79}, 155320 (2009).
\bibitem{gatti_onida} M. Gatti and G. Onida, Phys. Rev. B {\bf 72}, 045442 (2005).
\bibitem{siesta1} P. Ordej\'{o}n, E. Artacho, J. M. Soler, Phys. Rev. B (Rapid Comm.) {\bf 53}, R10441 (1996).
\bibitem{siesta2} J. M. Soler, E. Artacho, J. D. Gale, A. Garc\'{\i}a, J. Junquera, P. Ordej\'{o}n, D. S\'{a}nchez-Portal, J. Phys.: Condens. Matt. {\bf 14}, 2745 (2002).
\bibitem{espresso} P. Giannozzi et al., J. Phys.: Condens. Matt. {\bf 21}, 395502 (2009).
\bibitem{adler} S.L. Adler Phys. Rev. {\bf 126}, 413 (1962)
\bibitem{louie-chelikowsky-cohen} S. G. Louie, J. R. Chelikowsky, M. L. Cohen, Phys. Rev. Lett. {\bf 34}, 155 (1975).
\bibitem{LFglass} This condition has been verified with our code on a 192 atoms amorphous SiO$_2$.
\bibitem{pssb} Preliminary results for the Si$_{32}$/SiO$_2$ case were presented in Phys. Status Solidi B {\bf 247}, 2113 (2010).
\bibitem{Bruneval2005} F. Bruneval, S. Botti, and L. Reining, Phys. Rev. Lett. {\bf 94}, 219701 (2005).
\bibitem{Trani2007} F. Trani, D. Ninno, G. Iadonisi, Phys. Rev. B {\bf 75}, 033312 (2007).
\bibitem{bulutay2007} C. Bulutay, Phys. Rev. B {\bf 76}, 205321 (2007).
\bibitem{mirabella} S. Cosentino, S. Mirabella, M. Miritello, G. Nicotra, R. Lo Savio, F. Simone, C. Spinella, A. Terrasi, Nanoscale Res. Lett. {\bf 6}, 135 (2011).
\bibitem{gallas} B. Gallas, I. Stenger, C.-C. Kao, S. Fisson, G. Vuye, J. Rivory, Phys. Rev. B {\bf 72}, 155319 (2005).
\end{thebibliography}
\end{document}